\documentclass{mn2e}

\newcommand{\msun}{{\rm M}_{\sun}}

\usepackage[dvips]{graphicx}

\title{Jet radio emission in Cygnus~X-1 and its orbital modulation}

\author[A. Szostek and A. A. Zdziarski]
{
Anna Szostek\thanks{E-mail: asz@camk.edu.pl, aaz@camk.edu.pl}
and Andrzej A.~Zdziarski\footnotemark[1]\\
Centrum Astronomiczne im.\ M. Kopernika, Bartycka 18, 00-716 Warszawa, Poland\\
}

\date{Accepted . Received ; in original form }

\pagerange{\pageref{firstpage}--\pageref{lastpage}}
\pubyear{2006}

\begin{document}

\maketitle

\label{firstpage}

\begin{abstract}
We present results of our detailed theoretical study of the observed orbital
modulation of the radio emission in Cyg X-1. The modulation occurs due to
free-free absorption in the wind from the companion star varying with the
orbital phase, and our results put strong constraints on the spatial
distribution of the jet radio emission at the frequencies of 2--15 GHz. A
crucial role in enhancing the asymmetry of the wind absorption suffered by the
jet emission is played by the irradiation by X-rays emitted in the vicinity of
the black hole. This increases the wind temperature by more than order of
magnitude with respect to that of the wind of an isolated supergiant. The
observed phase lags of the minima of the radio emission with respect to the
spectroscopic zero phase strongly imply the bulk of the mass of the jet is
nonrelativistic ($\sim\! 5 \times 10^8$ cm s$^{-1}$) within the jet core. The
jet can, however, become relativistic outside the core. Also, the jet can have
a two-component structure, being slow on the outside and fast inside, in which
case its synchrotron-emitting part may be relativistic already in the core. We
also consider the observed superorbital modulation of the radio emission (with
the period of $\sim\! 150$ d) and find it can be explained by a jet precession
both causing variable wind absorption and changing the jet Doppler factor.
Finally, we consider the case of Cyg X-3, and show that its lack of observable
orbital radio modulation (in spite of strong modulation of X-rays) is explained
by that system being both much more compact and much more luminous than Cyg
X-1. 
\end{abstract}


\section{Introduction}
\label{intro}

Cyg X-1 is a high-mass X-ray binary with the orbital period of $P=5.6$ d (Gies
\& Bolton 1982), containing a black hole (Bolton 1972; Webster \& Murdin 1972)
and the O9.7 Iab supergiant HDE~226868 primary (Walborn 1973). Although the
primary nearly fills the Roche lobe, accretion is via a stellar wind (Gies \&
Bolton 1986a; Gies et al.\ 2003). Observations of optical lines indicate that
the wind departs from the spherical symmetry by being focused towards the
compact object. Still, only a few percent of the wind is accreted.

The orbital period has been detected in all energy bands where the source was
detected. X-ray modulation in the hard spectral state has the form of broad
intensity dips accompanied by spectral hardening centred on the superior
conjunction of the X-ray source, i.e., when it is behind the primary (e.g.,
Priedhorsky, Brandt \& Lund 1995; Wen et al.\ 1999; Lachowicz et al.\ 2006,
hereafter L06). The dip is a result of bound-free absorption by the wind (Wen
et al.\ 1999). Periodic changes in equivalent widths of UV absorption lines are
due to the orbital motion of a highly ionized, X-ray irradiated, region of the
stellar wind. In the superior conjunction, the equivalent widths achieve the
maximal values (Treves et al.\ 1980). In the visual band, the orbital period is
seen in radial velocity variations of absorption and emission lines (e.g.,
Bolton 1975) and in ellipsoidal variations (Walker 1972). 

The radio emission of Cyg X-1 is most likely due to the synchrotron process in
a jet, whose velocity has been claimed to be mildly relativistic, $\sim\! 0.6c$
(Stirling et al.\ 2001, hereafter S01; Gleissner et al.\ 2004). It has a flat
(in the energy flux, $F_\nu$) spectrum at $F_\nu\sim 10$--20 mJy in the X-ray
hard spectral state (see, e.g., Zdziarski et al.\ 2002 for a review of the
X-ray states of Cyg X-1), with no detection of either low or high frequency
cut-off between 2.2 and 220 GHz (Fender et al.\ 2000). 

Orbital periodicity of the radio emission was discovered by Han (1993).
However, more insight in it was possible thanks to long term radio monitoring
of Cyg X-1 by the Ryle Telescope at the Mullard Radio Astronomy Observatory at
15 GHz and the Green Bank Interferometer at 2.25 and 8.3 GHz (Pooley, Fender \&
Brocksopp 1999; L06). It was found that the total modulation depth {\it in the
hard state\/} increases with the frequency, from $\sim$0.03 at 2.25 GHz to
$\sim$0.3 at 15 GHz, and that there is a time lag of the minimum of the
modulation with respect to the superior conjunction decreasing with the
frequency, from $\sim$0.3 of the orbital period at 2.25 GHz to $\sim$0.15 of it
at 15 GHz. In the soft state, the orbital modulation takes place as well (L06),
but constraints on it are much weaker, due to the much lower on average but
flaring flux (Fender et al.\ 2006; L06).

We would like to point out that an orbital periodicity of radio emission is a
rare phenomenon in X-ray binaries. It appears to be seen in only two other
systems, LSI~$+61\degr 303$ (Taylor \& Gregory 1982) and Cir X-1 (Whelan et
al.\ 1977). Both of them are of rather different nature than Cyg X-1. Although
their compact objects orbit around stars undergoing heavy mass loss (as in Cyg
X-1), the orbits are wide and highly eccentric. During the periastron passage,
the accretion rate onto the compact object is greatly enhanced and both the
X-ray and radio fluxes increase (e.g., Coe et al.\ 1983; Marti \& Paredes 1995;
Haynes, Lerche \& Murdin 1980). On the other hand, this is certainly not the
case in Cyg X-1, in which the orbit is circular (Brocksopp et al.\ 1999b), and
the orbital separation is similar to the diameter of the primary. Thus, the
nature of the orbital modulation of its radio emission has to be different.

\begin{figure}
\centerline{\includegraphics[width=75mm]{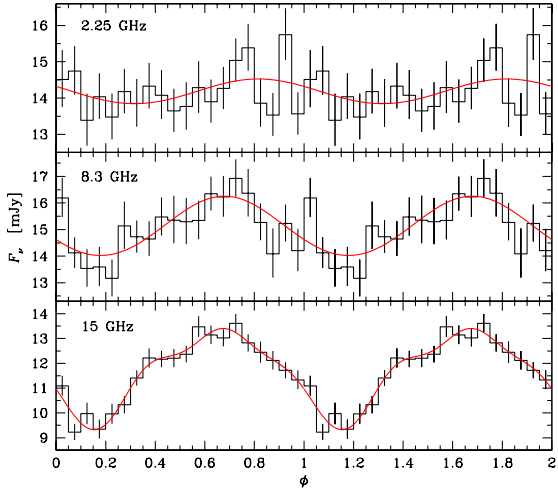}}
\caption{The average hard-state orbital modulation profiles at 2.25, 8.3 and 15
GHz (L06). The curves give the best fits with one (2.25, 8.3 GHz) and three (15
GHz) harmonics (L06). Two orbital periods are shown for clarity. } 
\label{data}
\end{figure}

The process likely to be responsible for the modulation of the radio emission
has been identified by Brocksopp, Fender \& Pooley (2002, hereafter BFP02).
They proposed it is free-free absorption in the stellar wind of the primary. It
is then analogous to the X-ray modulation, except for the absorption being
free-free rather than bound-free. 

Unfortunately, although the work of BFP02 generally explains the origin of the
radio modulation, it is plagued by a number of errors (as noticed by L06). Due
to those errors, their results are generally incorrect. For example, errors in
the equations and the numerical code used lead to the result that the
fractional depth of the modulation {\it decreases\/} with the increasing
inclination. This result is rather counterintuitive, as the system is fully
symmetric for a face-on observer and most asymmetric for an edge-on one, and it
is indeed incorrect. Then, the absorption in the wind outside the cylinder
defined by the orbit was underestimated by $\sim$40 orders of magnitude.
Furthermore, they assumed a uniform and rather low wind temperature, of 5000 K.
In reality, the wind is Compton-heated on the side of the X-ray source to
temperatures about two orders of magnitude higher (strongly reducing the
free-free opacity with respect to the treatment of BFP02). Also, the jet model
of BFP02 was relatively simplified, assuming that emission at a given frequency
is emitted entirely at a single point along a linear jet, and it did not
explain the phase lags.

In the present work, we consider models of the jet including its radial
emission profile as well as its bending. We consider only the hard-state data.
In modelling the stellar wind, we take into account its irradiation by the UV
and X-ray photons, and propagation during the orbital motion of the primary. We
model both the observed depth of the modulation as a function of frequency and
the corresponding lags of the minima of the folded radio lightcurves with
respect to the spectroscopic superior conjunction. This yields constraints on
the jet velocity and emission profiles of the radio emission.

\begin{table}
\centering
\caption{The average hard-state modulation depth and the phase shift [defined
in equation (\ref{ModDepth}) and calculated using the fitted model fluxes, see
Fig.\ \ref{data}], from L06. The uncertainties are $1\sigma$. }
\begin{tabular}{llll}
\hline
                       &      2.25 GHz       & 8.3 GHz          & 15 GHz        \\
\hline
$M_\nu$         & $0.045\pm 0.027$  & $0.138\pm 0.024$ & $0.304\pm 0.018$  \\
$\phi_{\nu,\rm min}$   & $0.316\pm 0.096$  & $0.179\pm 0.034$ & $0.153\pm 0.022$ \\
\hline
\end{tabular}
\label{Values}
\end{table}

\section{The observed modulation}
\label{observed}

We first present the main observational results on the orbital variability of
the radio emission from Cyg X-1, in order for this work to be self-contained.
Fig.\ \ref{data} shows the folded and averaged lightcurves from the Ryle and
Green Bank radio telescopes in the hard state (L06). The 15 GHz lightcurve has
the best statistical quality as well the highest modulation depth. As found in
L06, its fitting requires three harmonics. On the other hand, it is sufficient
to use a single sinusoidal dependence at 2.25 GHz and 8.3 GHz. 

The fractional modulation depth is defined as in L06,
\begin{equation}
M_\nu \equiv {F_\nu(\phi_{\nu,\rm max}) - F_\nu(\phi_{\nu,\rm min}) \over F_\nu(\phi_{\nu,\rm max})}, 
\label{ModDepth}
\end{equation}
where $F_\nu(\phi_{\nu,\rm max})$ and $F_\nu(\phi_{\nu,\rm min})$ are the
maximum and minimum fluxes in the averaged and folded lightcurves, observed at
the phases of $\phi_{\nu,\rm max}$ and $\phi_{\nu,\rm min}$, respectively, and
the phase lags, given by $\phi_{\nu,\rm min}$. Here, $\phi=0$ corresponds to
the spectroscopic superior conjunction of the X-ray source, and $\phi$ is
within the 0--1 interval. Table \ref{Values} gives the values of $M_\nu$ and
$\phi_{\nu,\rm min}$ of L06. They constrain the parameters of the binary and
the jet. 

\begin{figure}
\centerline{\includegraphics[width=83mm]{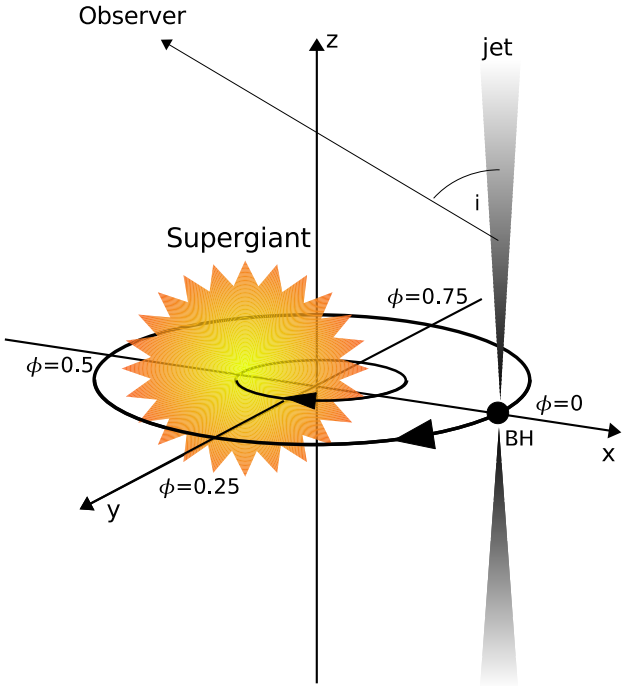}}
\caption{A schematic representation of Cygnus X-1 in the phase $\phi = 0$, with
a fast jet. (Note that the counter-jet has not yet been observed.) The centre
of mass is located within the primary. The observer is in the $xz$ plane. }
\label{system}
\end{figure}

\section{The model}
\label{model}

\subsection{The binary}
\label{s:system}

A schematic picture of the binary including the jet is shown in Fig.\
\ref{system}. The centre of our coordinate frame is in the centre of mass, and
the orbit is in the $xy$ plane. The observer is in the $xz$ plane. We note that
if the the jet is precessing (which may explain the observed superorbital
modulation, see Section \ref{super} below), the jet axis will be inclined with
respect to the binary plane. In our modelling of the orbital modulation below,
we neglect this effect since the precession amplitude is only poorly
constrained (e.g., L06) and the superorbital modulation may in principle be
alternatively explained by an intrinsic cycle of the supergiant wind.

The inclination, $i$, remains relatively uncertain. On the basis of analysis of
absorption lines, Gies \& Bolton (1986a) estimated $i = 33\degr \pm 5\degr$. On
the other hand, the polarimetric measurements of Dolan \& Tapia (1989) yield
$25\degr$--$67\degr$. In our study, we consider the inclination range of
30$\degr$--60$\degr$. 

We assume the black hole and the primary masses of $20\msun$ and $40\msun$
(Zi\'o{\l}kowski 2005), respectively, and a circular orbit. The radius of the
primary is taken as $r_{\star} = 1.58 \times 10^{12}$ cm (Gies et al.\ 2003).
Then, the separation corresponds to $a=2.28 r_{\star}=3.60\times 10^{12}$ cm,
and the orbital velocity is $3.13 \times 10^7$ cm s$^{-1}$ and $1.56\times
10^7$ cm s$^{-1}$ for the black hole and the primary, respectively. We
hereafter use $a$ as the unit, as it gives a measure of the asymmetry in the
photon paths during the orbital motion. The distance is assumed to be $D = 2$
kpc (Zi\'o{\l}kowski 2005).

\subsection{The wind}
\label{wind}

\begin{figure}
\centerline{\includegraphics[width=83mm]{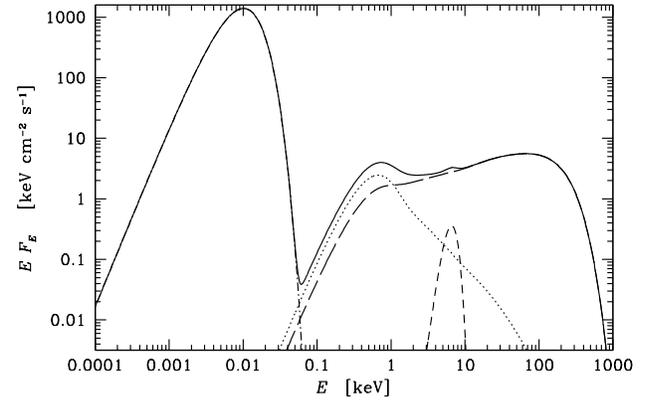}}
\caption{A characteristic intrinsic spectrum of Cyg X-1/HDE 226868 composed of
the UV blackbody from the primary at $T=3\times 10^4$ K (the dot-dashed curve) and a
hard state X-ray spectrum from the vicinity of the black hole. The X-ray
spectrum is from Frontera et al.\ (2001). The dots, long dashes and short
dashes correspond to the soft X-ray excess, the main hard X-ray Comptonization
component, and the Fe K$\alpha$ line, respectively. }
\label{spectrum}
\end{figure}

We assume the wind to be smooth, spherically symmetric and radiatively driven.
We follow Schaerer (1996) and Santolaya-Rey, Puls \& Herrero (1997) in assuming
a simple wind stratification at its base. The inner part, which we do not
consider here, is the stellar atmosphere. The outer part, i.e., the proper
wind, is parametrized by the velocity law, 
\begin{equation}
v(r) = v_{\infty}\left(1-{r_0 \over r}\right)^{\beta},\qquad r\geq r_\star,
\label{velocity}
\end{equation}
where $v_{\infty}$ is the terminal wind velocity,  
\begin{equation}
r_0 = r_{\star} \left[ 1-\left( {v_{\star} \over v_{\infty}}\right)^{1/\beta} \right] ,
\label{r0}
\end{equation}
and the lower boundary of the wind at $r_{\star}$ is chosen to correspond to
the isothermal sound speed, 
\begin{equation}
v_{\star}  \equiv v(r_{\star}) = \sqrt{{kT_{\star} \over \mu m_{\rm H}}},
\end{equation}
where $k$ is the Boltzmann constant, $m_{\rm H}$ is the hydrogen atom mass,
$\mu$ is the mean molecular weight, and $T_{\star}$ is the temperature of the
atmosphere, respectively. The particle number density at $r$ is related to the
mass-loss rate, $\dot M$, and and $v(r)$ via the continuity equation,
\begin{equation}
n(r) = {-\dot M \over 4 \pi m_{\rm H}\mu r^2 v(r) }.
\label{n_r}
\end{equation}
Note that equations (\ref{velocity}--\ref{n_r}) apply to a star at rest. Below,
we apply those equations taking into account the orbital motion, with $r -
r_\star$ measuring the distance travelled by a wind element (see below for
details). We adopt $-\dot M = 2.6 \times 10^{-6}\msun$ yr$^{-1}$ (Gies et al.\
2003), $v_{\infty} = 1.58 \times 10^8$ cm s$^{-1}$, $\beta=1.05$ (Gies \&
Bolton 1986b), and $T_{\star} = T_{\rm eff}$ where $T_{\rm eff}$ is the
effective temperature taken as $3\times 10^4$ K (Zi\'o{\l}kowski 2005). This
temperature at the assumed stellar radius yields $L_\star \simeq 1.5\times
10^{39}$ erg s$^{-1}$. For calculating $\mu$, we assume the wind to be ionized
with the abundances for an O9.7 Iab star (Lamers \& Leitherer 1993), which
yields $\mu=0.68$. We allow the wind to extend out to the infinity, which
assumption has a negligible effect on its absorption. 

We note that, with the exception of a very dense wind region close to the
stellar surface, the wind is optically thin for both X-ray and UV photons. The
resulting optically-thin approximation allows us to greatly simplify
calculations of the wind structure. In this approximation, the radiation field
at a distance $R$ away from a source of the luminosity, $L$, is determined only
by the geometrical dilution, $L/R^2$, with no radiative transfer calculations
being necessary. Moreover, at each point in the cloud with the number density,
$n$, the state of the gas for a given spectral shape depends solely on the
ionization parameter, 
\begin{equation}
\xi \equiv {L\over n R^2}
\label{xi}
\end{equation}
(Tarter, Tucker \& Salpeter 1969). Thus, we can apply the results of a single
model to a number of conditions with the same value of $\xi$. Based on the
results of Kallman \& McCray (1982), we apply the optically-thin approximation
for $\xi> 10^{2.5}\, {\rm erg\, cm\, s}^{-1}$. 

\begin{figure}
\centerline{\includegraphics[width=83mm]{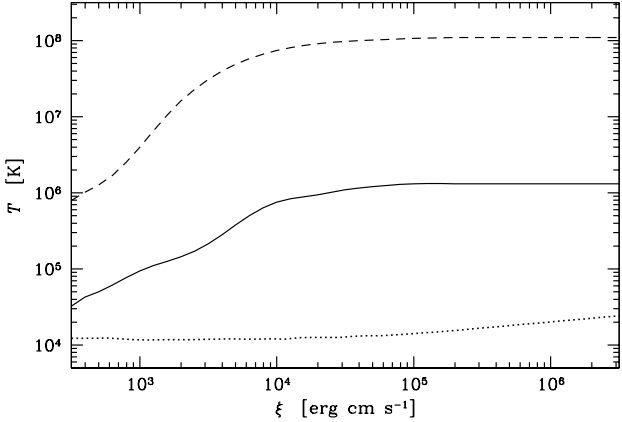}}
\caption{The gas temperature as a function of the ionization parameter
calculated using {\sc xstar}. We show the results for three irradiating
spectra: the composite X-ray/UV spectrum of Fig.\ \ref{spectrum} (solid curve),
the stellar $3 \times 10^4$ K blackbody (dotted curve), and the hard-state
X-ray spectrum (dashed curve). } 
\label{xitemp}
\end{figure}

In order to calculate the ionization and temperature structure of the wind, we
use the {\sc xstar} photoionization code (Bautista \& Kallman 2001). It
calculates the state of the gas surrounding an ionizing radiation source. The
gas is assumed to have a constant density but using the optically thin scaling
law allows us to apply the results to the optically-thin part of the wind.

The illuminating source is assumed to have the spectrum equal to the sum of the
hard-state spectrum of Cyg X-1 as modelled by Frontera et al.\ (2001) and the
primary blackbody spectrum of $T_\star = 3 \times 10^4$ K, shown in Fig.\
\ref{spectrum}. Our model spectrum is relatively similar to that of the model
\#5 of Kallman \& McCray (1982). We stress that in spite of the primary having
the luminosity about two orders of magnitude higher than that of the X-ray
source, it is necessary to include both sources in the calculations. In
particular, the gas temperature reaches the Compton temperature, $T_{\rm C}$,
in the limit of full ionization, 
\begin{equation}
T_{\rm C} = {\int E F_E {\rm d}E \over 4k \int F_E {\rm d} E}
\label{T_C}
\end{equation}
(e.g., Begelman, McKee \& Shields 1983), corresponding to the balance of the
Compton gains and losses only. Although the denominator is almost completely
dominated by the blackbody emission, the dominant contribution to the numerator
is from near the high-energy cutoff of the X-ray spectrum (in general, it is
the case for any spectrum harder than $F_E \propto E^{-2}$). The Compton
temperature of the X-ray component alone is as high as $\sim\! 10^8$ K (see
Fig.\ \ref{xitemp}). Then, the addition of the stellar blackbody increases the
Compton cooling [the denominator of equation (\ref{T_C})] and reduces $T_{\rm
C}$ approximately by the ratio between the two luminosities, i.e., by two
orders of magnitude, to $\sim\! 10^6$ K.

\begin{figure}
\centerline{\includegraphics[width=83mm]{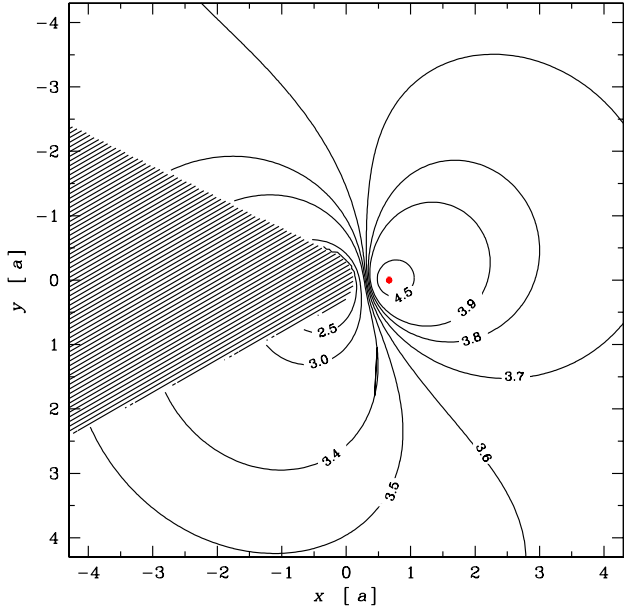}}
\caption{The distribution of $\log \xi$ in the plane of the orbit (the phase of
$\phi=0$ corresponds to the observer on the left-hand side). The corresponding
values of $T$ are given by the solid curve in Fig.\ \ref{xitemp}. The length
unit is the separation, $a$. The hatched area covers the primary and the wind
shadowed by it from the influence of the ionizing source. The small dot
represents the position of the black hole. The slight asymmetry of the contours
with respect to the line joining the star centres results from their
(clockwise) orbital motion and the finite wind velocity. }
\label{ionmap}
\end{figure}

The stellar blackbody emission has $L_\star\simeq 1.5\times 10^{39}$ erg
s$^{-1}$, whereas the X-ray luminosity is only $L_{\rm X}\simeq 2 \times
10^{37}$ erg s$^{-1}$ (see also the data compilation in Zdziarski et al.\
2002). We note that when calculating the ionization parameter, $\xi$ [equation
(\ref{xi})], {\sc xstar} uses only the luminosity in the 13.6 eV--13.6 keV
range, equal to $3.2\times 10^{38}$ erg s$^{-1}$ in our case. However, the full
spectrum is used in calculations of the temperature structure and the
distribution of different ions. The dependencies of $T(\xi)$ for the entire
spectrum of Fig.\ \ref{spectrum} and its two components separately are shown in
Fig.\ \ref{xitemp}.

We assume that the entire radiation source is located at the position of the
black hole (and thus $R$ is the distance from the black-hole center). This
approximation allows us to use {\sc xstar} with a single input spectrum of
Fig.\ \ref{spectrum} and for a range of the gas density. Although it introduces
relatively large errors close to the binary plane, it reproduces well the
actual temperature of the gas away from the plane, which region is important
for calculating absorption of the jet emission. 

Note that the gas density in the the inertial frame changes with the binary
motion. In this frame, the wind density at a given point has to be calculated
taking into account both the past motion of the primary and the motion of the
wind. However, we neglect the effects of the orbital motion and rotation on the
initial wind velocity. We also neglect the dynamics of the wind, e.g., effects
of gravity and radiation.

We then calculate the wind ionization structure, and the resulting temperature
distribution. The distribution of $\log \xi$ in the orbital plane is shown in
Fig.\ \ref{ionmap}. Because of a high density close to the surface of the
primary, the optically-thin approximation breaks down there. However, this
region does not contribute to absorption of the radio emission, which
originates at relatively large distances from the primary. Though the region
shadowed by the primary does not contribute to that absorption, we also
calculate its temperature, based the dotted curve in Fig.\ \ref{xitemp}. In the
vicinity of the black hole, $\log \xi \ga 4.5$, the wind is completely ionized
and its temperature equals $T_{\rm C}$. Note that since the wind reaches its
terminal velocity far away from the orbital plane, $n(r)\simeq n(R)\propto
R^{-2}$. There, $\xi$ reaches a constant, $\log \xi \simeq 3.6$ in our case,
which corresponds to $T \simeq 3.2 \times 10^5$ K. 

\begin{figure}
\centerline{\includegraphics[width=85mm]{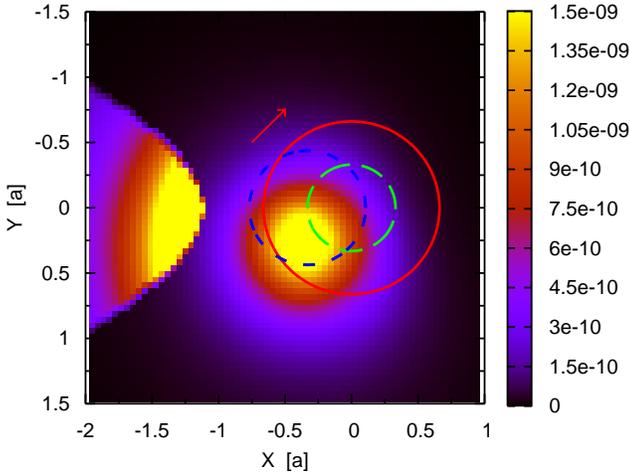}}
\caption{A map of the free-free absorption coefficient at 15 GHz at the height
of $2r_{\star}$ above the orbital plane. The phase of $\phi = 0$ corresponds to
the observer on the left-hand side. The image of the primary lags by $43\degr$
with respect to its actual position, shown by the short-dash circle, due to the
stellar motion and the finite wind velocity. The solid and long-dash circles
give the orbits of the black hole and the primary, respectively. The arrow
shows the direction of the orbital motion. The wind shadowed from the X-ray
irradiation by the primary is seen on the left.} \label{AbsorptionMap}
\end{figure}

The free-free absorption coefficient (in cgs units) is (Rybicki \& Lightman
1979), 
\begin{equation}
\alpha_{\nu} = 0.018 T^{-3/2} Z^2 n_{\rm e} n_{\rm i} \nu^{-2} \bar g_{\rm ff}\ {\rm cm}^{-1},
\label{alpha}
\end{equation}
where $n_{\rm e}$ and $n_{\rm i}$ is the number density of electrons and ions,
respectively. The velocity averaged Gaunt factor, $\bar g_{\rm ff}$, is
approximated as
\begin{equation}
\bar g_{\rm ff} = 9.77  + 1.27 \log {T^{3/2} \over Z \nu}
\label{gaunt}
\end{equation}
(Leitherer et al.\ 1995). For simplicity, we hereafter use $Z = 1$. Combining
equations (\ref{velocity}--\ref{r0}), (\ref{n_r}) and
(\ref{alpha}--\ref{gaunt}), and using the numerical values assumed above, we
obtain, 
\begin{equation}
\alpha_{\nu}(r) = {4.38 \times 10^{67} \bar g_{\rm ff}\over T^{3/2} r^{4} \nu^{2} }
\left(1-{1.56 \times 10^{12}\over r}\right)^{\!-2.1}\! {\rm cm}^{-1}.
\label{alpha2}
\end{equation}
In equations (\ref{alpha}-\ref{alpha2}), $T$ is in K, $r$ is in cm and $\nu$ in
Hz. We integrate the absorption coefficient along the line of sight from the
place of radio emission at each phase to the observer, which gives us the
optical depth and, in turn, the observed flux, $F_{\nu}$. 

As a result of the stellar motion and the finite wind velocity, the image of
the primary reflected in the wind density, or the absorption coefficient, is
phase-lagged with respect to the actual position of the primary at that time.
We illustrate this effect in Fig.\ \ref{AbsorptionMap}. It shows the absorption
coefficient at the height of $z=2r_\star$ above the orbital plane. We see a
shift of the stellar image with respect to the actual primary position. The
effect is visible here because the velocity of the wind is larger than the
orbital velocity by only one order of magnitude, with the wind travelling the
distance of $\sim 17.5 a$ away from the stellar surface during one orbital
period. This phase lag of the primary increases with the height above the
orbital plane, and will by itself cause a phase shift in the orbital modulation
of the radio emission with respect to the actual ephemeris. We take it into
account in our calculations; however, we find it cannot by itself explain the
observed phase lags. We also see the shadow wind in Fig.\ \ref{AbsorptionMap}.
Because of its relatively low temperature, it has higher free-free optical
depth than the rest of the wind.

\subsection{The jet}
\label{jet}

\begin{figure}
\centerline{\includegraphics[width=85mm]{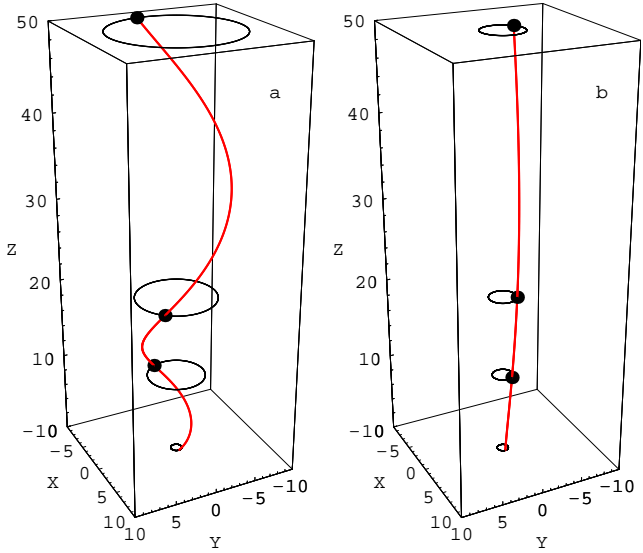}}
\caption{Two jets with $v_0 = 0$ and a low and a high acceleration, (a) $5
\times 10^2$ cm s$^{-2}$, (b) $5 \times 10^3$ cm s$^{-2}$, respectively. The
black hole orbit is shown in the orbital plane, and three circles above it show
the jet orbit at the height of $z= $ 10, 20 and $50 a$.} 
\label{CurvedJet}
\end{figure}

We assume that the radio emission originates in the jet. We neglect its
transverse size and consider only its longitudinal structure. We assume that
the geometry of the jet is directly defined by the initial condition at its
base, namely the initial vertical velocity, $v_0$, the acceleration, $a_0$, and
the perpendicular instantaneous orbital velocity, $2 \upi R_{\rm BH}/P$, where
$R_{\rm BH}$ is the orbital radius of the black hole. (Here, we do not take
into account a possible jet precession and thus $v_0$ and $a_0$ have only
vertical components.) We assume a continuous jet moving away from the place of
the ejection. The trajectory of a jet element has the parametric form, 
\begin{equation}
x(t) = R_{\rm BH} \cos{(2 \upi \phi)} - {2 \upi t \over P} R_{\rm BH} \sin{(2 \upi \phi)},
\end{equation}
\begin{equation}
y(t) = R_{\rm BH} \sin{(2 \upi \phi)} + {2 \upi t \over P} R_{\rm BH} \cos{(2 \upi \phi)},
\end{equation}
\begin{equation}
z(t) = v_0 t + {1 \over 2} a_0 t^2, 
\end{equation}
where $t$ is the time elapsed from the ejection. If the jet is fast, it remains
nearly perpendicular to the orbital plane, as shown schematically in Fig.\
\ref{system}. For a slower jet, the relative contribution from the orbital
velocity is important, and it bends, i.e., its position at a given radius lags
behind the black hole. The projection of the motion of a single blob in the jet
onto the orbital plane forms a spiral. Fig.\ \ref{CurvedJet} shows two examples
for $v_0 = 0$ and two different values of $a_0$. 

We then consider the jet emissivity profile. The simplest one corresponds to
discrete emission, where each point of the jet emits a (single) different
frequency (which approximation was used by BFP02). As a more realistic
parametrization of continuous jet emission, we use a double-power-law profile,
\begin{equation}
S_{\nu}(z') = {K_{\nu} (z'/\mu_\nu)^{n_{\nu}} \over 1 + (z'/{\mu}_{\nu})^{m_{\nu}}},
\label{double}
\end{equation}
where $S_{\nu}(z')$ is the radio flux emitted at $z'$ and $K_{\nu}$,
$\mu_{\nu}$, $n_{\nu}$ ($>0$) and $m_{\nu}$ ($>n_\nu$) are $\nu$-dependent
parameters. The variable $z'$ gives the distance along the jet from its base.
Hence, only in the case of the fast, straight, jet, $z'\approx z$.

If we consider only relative emission profiles, without specifying the absolute
value of the emitted flux, we can set $K_{\nu}$ to the value normalizing the
profile to unity at the maximum. The maximum flux is emitted at 
\begin{equation}
z'=\mu_{\nu} \left({m_{\nu} \over n_{\nu}}-1\right)^{-1/m_{\nu}},
\label{zmax}
\end{equation}
and
$S_\nu(z') \propto (z')^{n_\nu-m_\nu}$ and $\propto (z')^{n_\nu}$ far
above and far below the maximum, respectively. The profile normalized to unity
is given by 
\begin{equation}
S_{\nu}(z')= {m_\nu\over n_\nu} \left({m_\nu\over n_\nu}-1 \right)^
     {(n_\nu/m_\nu)-1} 
{(z'/\mu_\nu)^{n_\nu}\over 
     1 + (z'/\mu_\nu)^{m_\nu}} .
\label{double_norm}
\end{equation}

\section{Results}
\label{results}

\begin{figure}
\centerline{\includegraphics[width=85mm]{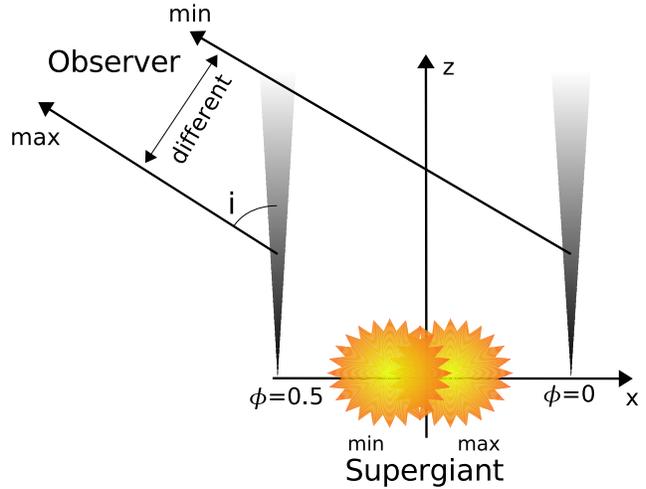}}
\caption{A schematic representation of the system, illustrating the difference
between the lines of sight at $\phi=0$ and 0.5. The optical paths at these two
phases differ from each other both inside and outside the cylinder defined by
the black hole orbit. }
\label{LineOfSight}
\end{figure}

Orbital modulation is produced when the source of the radio emission travels
through the wind and thus it is observed through different optical depths, as
illustrated in Fig.\ \ref{LineOfSight}. The depth of the modulation, its phase
shift and the average transmitted flux depend on the spatial profile of the
radio emission, the jet velocity profile, and the inclination, which
dependencies we investigate.

In general, as it is apparent from Fig.\ \ref{LineOfSight}, the difference
between the lines of sight is not confined to the space inside the cylinder
defined by the black hole orbit. Because of a strong dependence of the
absorption coefficient on the distance from the primary, $\alpha_{\nu} \propto
r^{-4}$, a small difference in the line of sight can strongly affect the
observed modulation. It is therefore not correct to neglect the differences in
the absorption outside the orbital cylinder (as it was done in BFP02). 

\subsection{Discrete jet emission}
\label{point}

\begin{figure}
\centerline{\includegraphics[width=85mm]{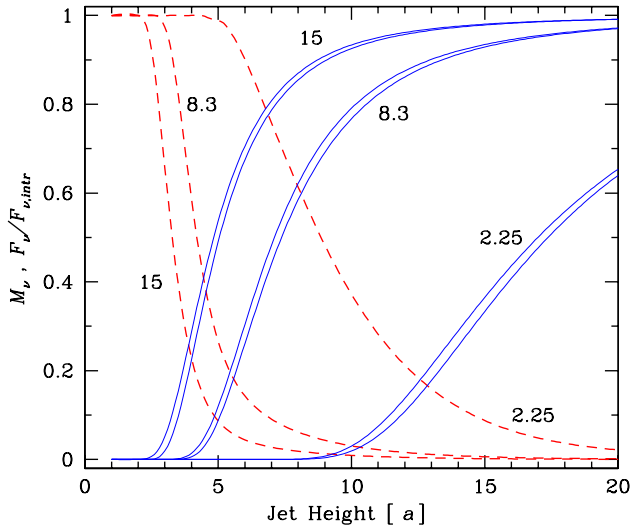}}
\caption{The modulation depth (dashed curves) and the fractional transmitted
flux (solid curves) at a given point along a jet seen at $i=40\degr$. The
transmitted fluxes are shown for both $\phi = 0$ (lower curve in pair) and
$0.5$. Each curve is marked with the corresponding frequency in GHz. The curves
take into account the binary motion. } 
\label{AbsorptionProfile}
\end{figure}

We first consider a simple model with a fast jet, perpendicular to the orbital
plane. We compare the observed emission from points along the jet at $\phi=0$
and 0.5. We show the results in Fig.\ \ref{AbsorptionProfile}. We see that the
strength of the modulation is close to unity near the base of the jet, and then
decreases fast with the height. However, the intrinsic jet emission is very
strongly attenuated in the regions where the jet modulation is close to unity.
Excluding these region, the fast decrease of the modulation strength with the
height implies that the spatial emissivity profile at a given frequency has to
be narrow in order to yield a substantial modulation. These considerations
point to the relative narrowness of the parameter space where strong radio
modulation by wind absorption can be achieved, and may explain why Cyg X-1 is
the only binary showing radio modulation due to that process. 

\begin{figure}
\centerline{\includegraphics[width=85mm]{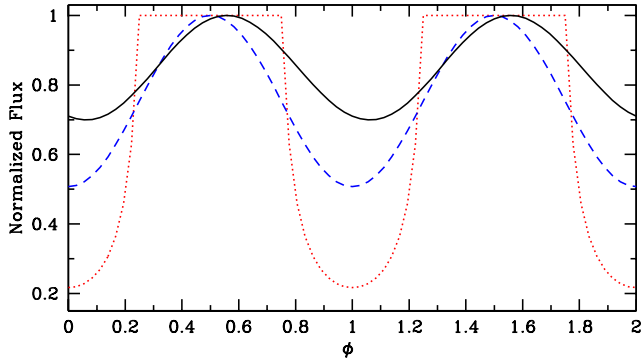}}
\caption{The normalized 15-GHz lightcurves for three different models. In all
cases, we assume emission from a point of a straight jet at $z=3.74 a$ and
$i=40\degr$. The dotted curve assumes the fixed primary and the wind absorption
model as in BFP02. The dashed curve is for the fixed companion but properly
accounting for the absorption. The solid curve is for our model, with the stars
orbiting around the centre of mass. The visible phase lag in this curve is due
to the phase lag of the image of the primary (see Fig.\ \ref{AbsorptionMap}),
and it is $\simeq 0.06$ (corresponding to $\simeq 21\degr$). }
\label{LightCurves}
\end{figure}

We then stress the importance of properly accounting for the system geometry,
see Fig.\ \ref{LightCurves}. We plot here three normalized lightcurves for
models with three different system geometries. First (dots), the position of
the primary is fixed at the centre with the black hole revolving around it on a
circular orbit, and the absorption is confined to the cylinder defined by the
black hole orbit, as assumed by BFP02. Second (dashes), we keep the former
assumption but take into account the entire absorption. Third (the solid
curve), we consider the actual case with the stars orbiting around the centre
of mass. In all cases, we assume the emission from a single point of the
straight jet. We see that the shape of the dotted curve disagrees with the
observed lightcurve, in particular, there is no absorption at
$\phi=0.25$--0.75. Then, accounting for the orbital motion of the both stars
(solid curve) results in shallower modulation as compared to the case with the
fixed companion. The phase shift of the minimum of the solid curve is due to
the apparent displacement of the companion, see Section \ref{wind} and Fig.\
\ref{AbsorptionMap}.

\begin{figure}
\centerline{\includegraphics[width=85mm]{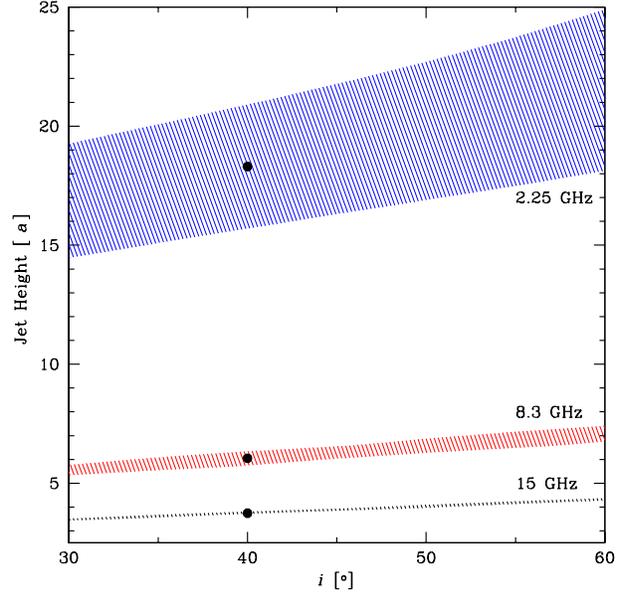}}
\caption{The shaded areas show the locations of the radio emission in the
$i$--$z$ space allowed by the observed modulation depths (Table \ref{Values}),
assuming discrete emission of a vertical jet. The black dots give the values
assumed in Fig.\ \ref{estimates} below. } 
\label{ModulationMap}
\end{figure}

The difference between the minimum and maximum optical depth, and thus the
modulation depth for emission from a single point, depends on the height above
the orbital plane, $z$, and the inclination, $i$. We show this relationship in
Fig.\ \ref{ModulationMap} for the observed modulation depths (Table
\ref{Values}). For example, $M_\nu \simeq 0.3$ at 15 GHz at $i=40\degr$
corresponds to point emitting at $z\simeq 3.74 a$. 

The allowed emission regions at different frequencies are distinctly different,
with lower frequencies emitted higher. For each $\nu$, the region below the
respective shaded area is occupied by points whose transmitted flux has higher
modulation depth with respect to observed ones. Emission from points above the
shaded area is modulated with lower amplitude. 

\begin{figure}
\centerline{\includegraphics[width=85mm]{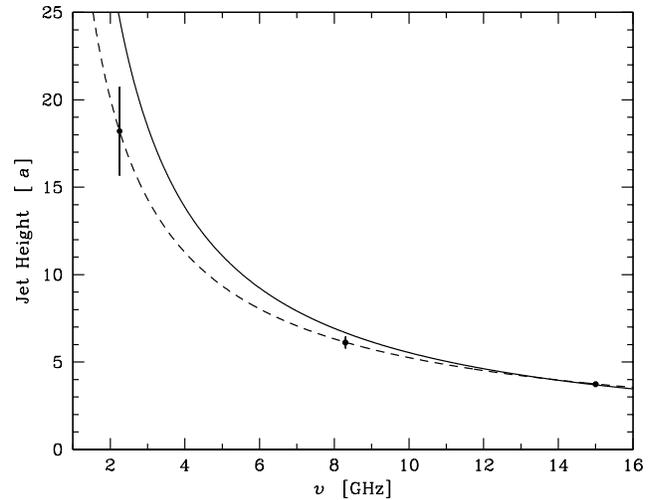}}
\caption{The three points shown in Fig.\ \ref{ModulationMap} with 1$\sigma$
error bars, fitted with the $z\propto 1/\nu$ theoretical relationship (solid
curve) and the best fit for $z\propto 1/\nu^{\eta}$ ($\eta=5/6$; dashed curve). }
\label{estimates}
\end{figure}

\begin{figure}
\centerline{\includegraphics[width=85mm]{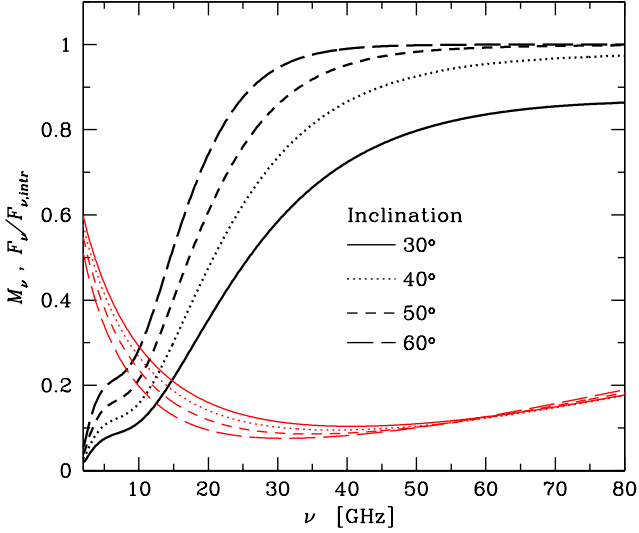}}
\caption{The relationships between $M_\nu$ and $\nu$ based on equation
(\ref{fit2}), for four values of $i$. The heavy and light curves correspond to
the modulation depth and the fractional transmitted flux, respectively.}
\label{Inclination}
\end{figure}

We then compare our estimated emission distributions with the $z \propto 1/\nu$
relationship obtained by Blandford \& K\"onigl (1979). This relation follows
from the position of the (photospheric) height where the jet becomes optically
thin for synchrotron self-absorption. The three points shown in Fig.\
\ref{ModulationMap} have been fitted (including the errors on $z$ estimated
from Fig.\ \ref{ModulationMap}) with $z = A/\nu$, where $A\propto L_{\rm
J}^{2/3}$, and $L_{\rm J}$ is the jet power in relativistic electrons and
magnetic fields (Blandford \& K\"onigl 1979). The best-fit relation, 
\begin{equation}
{z\over a} = {55.4\, {\rm GHz}\over \nu}, 
\label{fit}
\end{equation}
shown in Fig.\ \ref{estimates}, does not agree with our model (based on the
observed modulation depths) for 2.25 and 8.3 GHz within the 1$\sigma$ error
ranges. However, it does agree with the data within their 90 per cent
confidence intervals. We also plot the best-fit curve for a general power law
dependence, for which we find,
\begin{equation}
{z\over a} = {A \over (\nu/1\,{\rm GHz})^{\eta}}, \quad \eta \simeq 5/6,\quad A \simeq 35.8.
\label{fit2}
\end{equation} 
Thus, the best-fit power-law index is only slightly lower than that given by
Blandford \& K\"onigl (1979). The relationships at other inclinations than
$40\degr$ are slightly different, but generally the heights of the 2.25 GHz and
8.3 GHz emission is slightly less than that predicted by equation (\ref{fit}). 

We then obtained the positions and modulation depths for other frequencies
extrapolating the observed modulation depths using equation (\ref{fit2}), see
Fig.\ \ref{Inclination}. The curvature of the modulation depth at the
low-frequency end results from changes of the apparent orbital separation
related to the displacement of the primary image due to the motion of both
stars and the finite wind velocity (see Fig.\ \ref{AbsorptionMap}). At $\nu \ga
40$ GHz, $M_\nu \rightarrow 1$, i.e., the emission at the minimum is then very
strongly absorbed. The phase-averaged fractional transmitted flux above 40 GHz
is $\sim \! 0.1$. We also see that an increase of $i$ increases the modulation
depth.

\subsection{Double-power-law profile}
\label{double:2}

\begin{figure}
\centerline{\includegraphics[width=85mm]{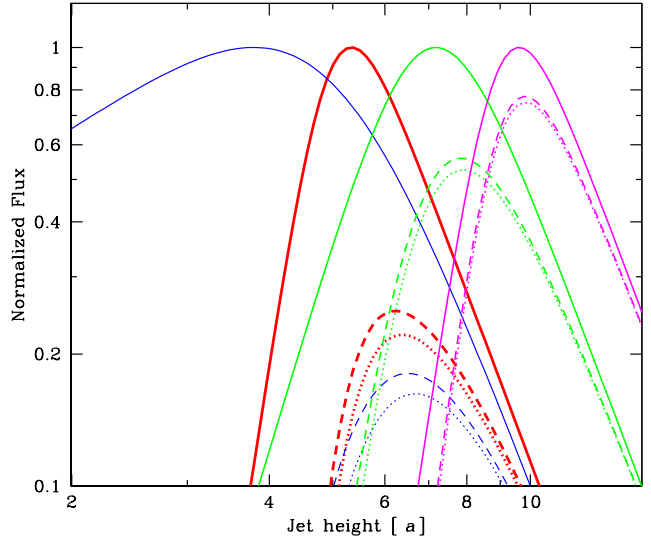}}
\caption{Examples of the intrinsic double-power-law profiles normalized to
unity at the maximum (solid curves), together with the transmitted profiles for
$\nu=8.3$ GHz, $i=40\degr$, and $\phi=0.5$ (dashed curves) and $\phi=0$ (dotted
curves). The red, blue, green and magenta curves correspond to ($\mu_{\nu}/a$,
$n_{\nu}$, $m_{\nu}$) equal to $(5, 10, 14)$, $(5, 1, 5)$, $(7, 5, 9)$,
$(9, 10, 14)$, respectively.}
\label{profiles}
\end{figure}

As an approximation to realistic jet emission, we consider double-power-law
profiles, equation (\ref{double}). This form of the spatial distribution can
closely reproduce detailed jet models, e.g., those of Hjellming \& Johnston
(1988) and Kaiser (2006). In Fig. \ref{profiles} we show examples of that
profile for a number of parameters together with the same profiles transmitted
through the wind at the phases of $\phi = 0$ and $0.5$. The difference between
the transmitted fluxes integrated over the jet length at those two phases gives
(approximately, as the true minima have some phase shifts) the modulation
depth. We can see that the higher the point of emission along the jet, the
lower the difference between the two transmitted profiles, and hence the lower
modulation. The low-$z$ ($\propto z^{n_{\nu}}$) branch of the profiles is
almost completely absorbed and thus the shape of the transmitted profiles and
the modulation depth is only weakly dependent on $n_{\nu}$. However, the
fractional transmitted flux (i.e, with respect to the unobserved intrinsic
emission) does significantly depend on $n_{\nu}$. 

We have then created contour maps in the location--width plane. The former is
represented by $\mu_{\nu}$, determining the profile peak [equation
(\ref{zmax})], and the latter, by $m_{\nu}-n_{\nu}$, which is approximately
[see equation (\ref{double_norm})], inversely proportional to its width. Fig.\
\ref{BrokenInf1} shows two maps corresponding to profiles with significant
($n_{\nu} = 1$) and small ($n_{\nu} = 10$) contributions to the intrinsic
emission from below the profile maximum. The hatched areas represent the
profiles with the observed modulation depth (Table \ref{Values}) at different
$\nu$ and $i$. As it can be inferred from Fig.\ \ref{BrokenInf1}, broader
profiles correspond to lower $z$ and higher $i$. Also profiles emitted at lower
$z$ have lower fractional transmitted fluxes. On the other hand, narrow
profiles have to be emitted at higher $z$, and have the value of $\mu_{\nu}$
approximately equal to the corresponding location of point emission profiles
shown in Fig.\ \ref{ModulationMap}. 

The areas of the parameter space allowed by the observations are smaller for
$n_\nu=10$ because these profiles are narrower, and the corresponding
modulation depth quickly decreases with the increasing height of the jet. These
profiles are also shifted towards lower values of $\mu_{\nu}$ than the
$n_{\nu}=1$ profiles. These effects are more prominent at lower frequencies. 

The phase shifts we are able to obtain for the double-power-law profile along a
straight jet are within about $\pm 0.08$ (due to the effects of the finite wind
velocity coupled with the orbital motion). They are not sufficient to explain
the observed phase lags (Table \ref{Values}).

\subsection{A slow jet}
\label{slow}

\begin{figure}
\centerline{\includegraphics[width=85mm]{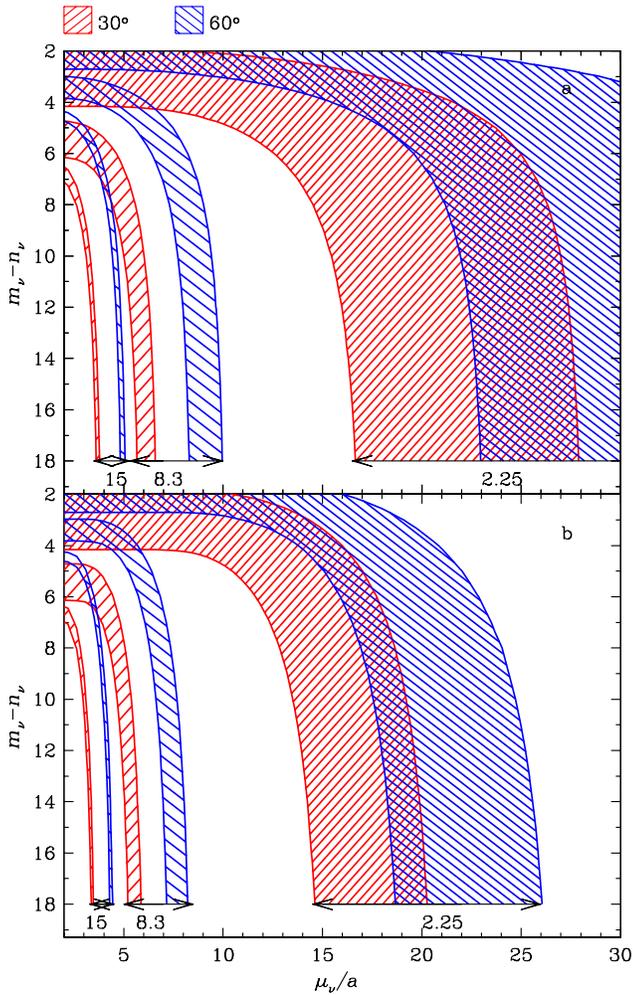}}
\caption{The regions in the space of the peak position, $\mu_{\nu}$, vs.\
$m_{\nu}-n_{\nu}$ of the double-power-law emissivity profile allowed by the
observed modulation depths (Table \ref{Values}), for different values of $\nu$
and $i$ and for (a) $n_{\nu}=1$ and (b) $n_{\nu}=10$. }
\label{BrokenInf1}
\end{figure}

Here, we consider the jet bending due to its initial (binary) motion together
with the black hole, see Section \ref{jet}. As it can be seen from Fig.\
\ref{CurvedJet}, a curved jet bends outside the cylinder defined by the black
hole motion. Thus, the amplitude of the optical depth variations is higher than
in the case of the vertical jet, which will then increase the strength of the
modulation. Also, the resulting lag behind the compact object will lead to a
lag in the phase of the maximum absorption (i.e., the minimum flux), as
observed. 

In general, the flux phase profile depends on the inclination, $i$, the profile
parameters ($\mu_\nu$, $m_\nu$, $n_\nu$), and on the velocity profile, given by
$v_0$ and $a_0$. Then, in order to reduce the parameter space, we consider here
only two velocity profiles, one with $v_0>0$ and $a_0=0$ (corresponding to
acceleration only at the jet base), and one with $v_0=0$ and a constant
$a_0>0$. In addition to the constraints from $M_\nu$, we now also take into
account the constraints from the phase lag of the minimum flux, $\phi_{\nu,{\rm
min}}$ (Table \ref{Values}). 

\begin{figure*}
\centerline{\includegraphics[width=14cm]{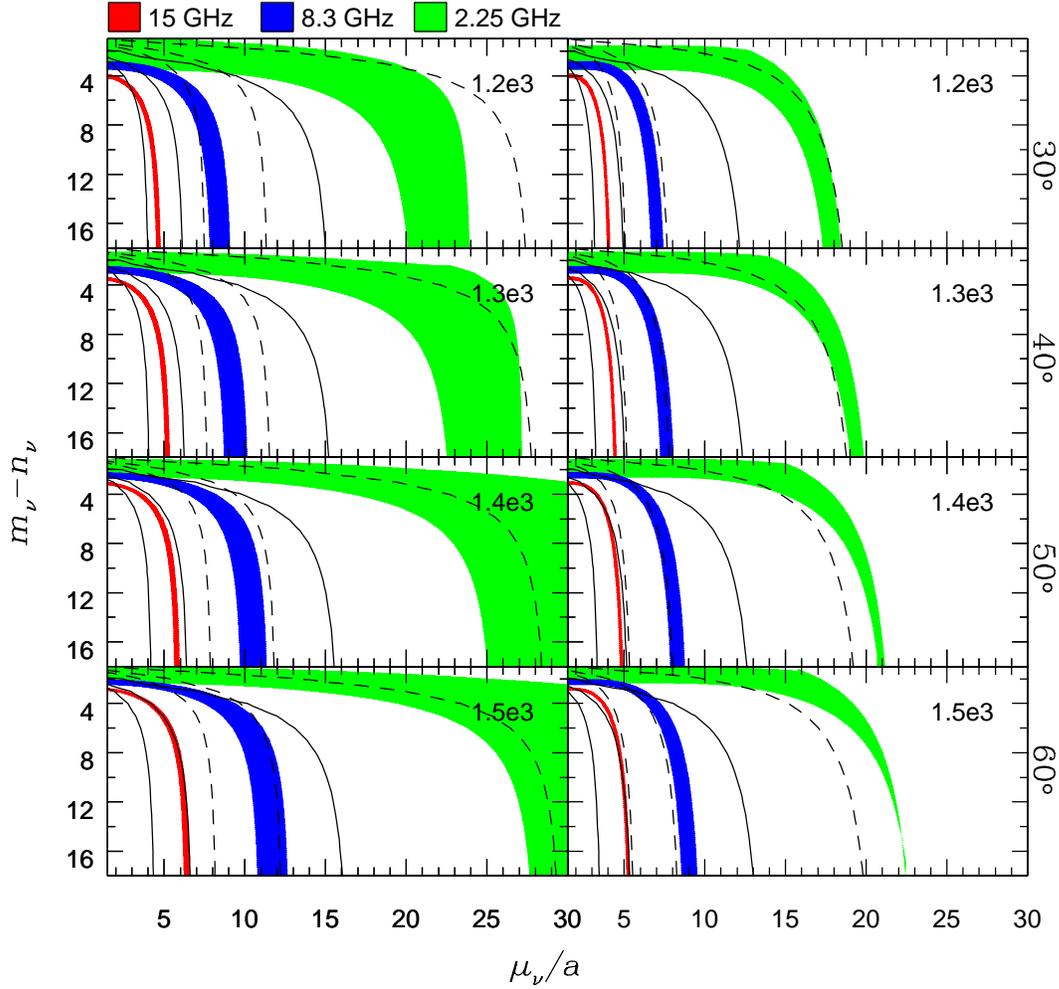}} 
\caption{Constraints on the double power-law emissivity profiles, with
$\mu_\nu$ giving the position of the peak emission, and $m_{\nu}-n_{\nu}$ its
width, for a jet with $v_0=0$ and constant acceleration (given on each panel in
units of cm s$^{-2}$). The left and right columns correspond to $n_{\nu} = 1$
and 10, respectively, and each row correspond to a different value of $i$
(marked on the right). The parameters allowed by both the observed modulation
depth and the phase lag are marked by the coloured areas separately for each
$\nu$. The solid and dashed curves correspond to the fractional transmitted
fluxes of 0.1 and 0.5, respectively, at $\nu= 15$ GHz, 8.3 GHz and 2.25 GHz,
from left to right. 
} 
\label{BrokenMap}
\end{figure*}

First, we find that {\it no\/} jet with $v_0>0$ and $a_0=0$ can satisfy the
observational constraints. At most, the observational constraints can be
satisfied at only one of the observed frequencies. Then, we consider the case
with $v_0=0$ and constant $a_0>0$, which we find can easily satisfy those
constraints. Fig.\ \ref{BrokenMap} shows the ranges of the parameters allowed
by the data for a selection of values of $a_0$. We show the results for
$n_{\nu}=1$ and 10, analogous to Fig.\ \ref{BrokenInf1} for the case of the
straight jet. The coloured regions are those where both of the constraints are
satisfied for a given frequency. Such regions exist for each of the observed
frequencies and given an appropriate choice of $a_0$ [cm s$^{-2}$] in the range
of $(1.1$--$1.4)\times 10^3$ for $i=30\degr$, $(1.1$--$1.5)\times 10^3$ for
$i=40\degr$, $(1.2$--$1.6)\times 10^3$ for $i=50\degr$ and $(1.2$--$1.7)\times
10^3$ for $i=60\degr$. The values of $a_0$ used in Fig.\ \ref{BrokenMap}
correspond approximately to the middle of these ranges. It is, of course,
possible that the jet acceleration is not constant, but as this is not required
by the data, we do not consider such models. We also show the corresponding
fractional transmitted fluxes. They are allowed to be rather large, typically
$\sim\! 0.2$--0.5 for broad profiles ($n_{\nu}=1$) and $\ga 0.5$ for narrow
profiles ($n_{\nu}=10$). 

The acceleration is relatively slow and the jet will be curved, with the
curvature somewhere in between those shown in Fig.\ \ref{CurvedJet}(a) and (b).
At the distance of $20 a$ ($7.2 \times 10^{13}$ cm), around which the 2.25 GHz
radiation is emitted in our model, a jet with $a_0=1.3 \times 10^3$ cm s$^{-2}$
will reach $v\simeq 4.3\times 10^8$ cm s$^{-1}$ during $3.3 \times 10^5$ s. At
this point, it will lag behind the black hole by $\sim\! 170\degr$ and it will
be at the distance from the orbital axis of 4.5 times the radius of the
black-hole orbit.

\section{Discussion}
\label{discussion}

\subsection{General features of the model}
\label{general}

There are a few essential features of our model, necessary to account for the
observed orbital modulation depths and the phase lags. One is the wind heating
due to the irradiation by the X-rays. This naturally creates a strong
temperature anisotropy, and consequently strong anisotropy of the free-free
absorption coefficient. Though the X-ray luminosity is $\sim\! 10^2$ times
lower than the UV luminosity of the primary, Compton heating is proportional to
the {\it product\/} of the irradiating energy flux and the photon energy. The
latter is $\sim\! 10^4$ times higher for the X-rays than for the UV, and
consequently, the X-ray heating is $\sim\! 10^2$ times higher than the UV
heating. We also point out that the range of the temperatures of the wind
obtained by us, $T\simeq 10^{4.5}$--$10^6$ K, is in full agreement with that
found from {\it Chandra\/} soft X-ray observations of Cyg X-1 by Miller et al.\
(2005). 

For test purposes, we have also run models with a uniform wind temperature
equal $10^4$ K. They required strong fine-tuning in order to achieve the
observed modulation depths, as well as  very low resulting transmitted fraction
of the intrinsic jet flux, $\la 10^{-2}$. Inclusion of X-ray heating resolves
those problems.

Another essential feature is the required low velocity of the bulk of the mass
of the jet, necessary for the jet bending required to account for the observed
phase lags of the flux minima (with respect to the superior conjunction of the
black hole). The velocity is nonrelativistic within the jet region constrained
by the data. The jet bending results then from the initial jet velocity in the
binary plane (due to the orbital motion of the black hole) being comparable to
its vertical velocity component. We have searched for other ways to account for
the observed phase lags, and have found none. In particular, the lags
associated with the finite wind velocity together with the motion of the
primary star cannot account for the observations. 

We note that the modulation depth increases with $\nu$ in the 2.25--15 GHz
range. Our model reproduces this increase. Not unexpectedly, it predicts the
depth of the orbital modulation to increase further at $\nu>15$ GHz. It would
be of high interest to test this prediction observationally. We note that there
have been rather few observations at $\nu>15$ GHz so far (Fender et al.\ 2000;
note some errors in the observation dates given there), insufficient for that
test.

A related issue is the shape of the {\it intrinsic\/} jet spectrum. The
observed average, absorbed, spectrum, is approximately given by $F_\nu\propto
\nu^\alpha$ with $\alpha\simeq 0$ in the $\sim\! 2$--200 GHz range (Fender et
al.\ 2000). The intrinsic spectrum of the jet is also likely to be a power law,
in general with a value of $\alpha$ allowed to be different than 0. However, if
the intrinsic index is substantially different from 0, then obtaining the
$\alpha\simeq 0$ power law would require a specific power-law absorption law.
We can obtain it with our model. However, given the relatively large number of
the free parameters compared to the relatively few observational constraints,
the resulting model constraints would be rather loose. On the other hand, the
most natural way to obtain the observed $\alpha\simeq 0$ is with the
transmission fraction through the wind being high, close to 1. As we show in
Fig.\ \ref{BrokenMap}, there is a relatively large parameter space with that
fraction being high, $\ga 0.3$, and thus this is indeed possible.

The radio maps obtained from observations at 8.4 GHz (S01) show the jet to form
an extended feature. The flux reaches the maximum at the core of an angular
size of $\simeq 3$ mas, and then it decreases over $\sim\! 12$ mas away from
it. For $i$ from $60\degr$ to $30\degr$, 3 and 15 mas correspond to (29--$50) a
\simeq (1.0$--$1.8) \times 10^{14}$ cm and (140--$250) a\simeq (5$--$9)\times
10^{14}$ cm, respectively. Our constraints are for the distances along the jet
of $z\la 30 a$ at 2.25 GHz and $z\la 15 a$ at 8.3 GHz. Thus, they apply only to
the core of the jet. 

Still, we can compare the predictions of our model with the radio maps of S01
at larger distances using the double-power-law profiles. We find that the
observed extended structure of the jet is substantially more luminous than the
predictions of our simple models. The propagation of the jet beyond the core
may be affected by various effects, e.g., its moving away from the stellar wind
region, and this issue is beyond the scope of the present work. 

We note that the 2.25--15 GHz monitoring data used by us are integrated over
the entire observed source. The emission at a given frequency from outside of
the jet core is very unlikely to be modulated by the stellar wind. Thus, the
fractional modulation of the emission close to the jet base has to be actually
somewhat higher than that used by us, in order to account for the constant part
emitted outside the core. According to the results of S01, the 8.4 GHz flux
from the extended part is less than or similar to that of the core. Then, the
fractional modulation of the core emission should be higher by a factor $\la 2$
than that averaged over the entire structure, but the values of the phase lags
are not affected. 

It is possible to reproduce the core emission profiles with such a modulation
depth in our model. The required profiles are located to the left of the 8.3
GHz region in Fig.\ \ref{BrokenMap}, which introduces only a small change in
the value of the acceleration, reducing it only by at most 10 per cent. Then,
the results of S01 show that the 15.4 GHz emission is mostly in the core of the
extend of $\la 2$ mas, with only $\la 10$ per cent of the emission outside the
core. This is approximately compatible with our spatial emission profiles.
Then, no radio maps are available at $\sim\! 2$ GHz. 

Our models yield the jet acceleration of $\sim\! (1.1$--$1.7)\times 10^3$ cm
s$^{-2}$. At $20 a$, the jet achieves the velocity of $\sim (4$--$5)\times
10^8$ cm s$^{-1}$, and $\sim\! 3$ times that at $200 a$, i.e., $\ll c$ in both
cases. On the other hand, S01 has estimated the jet velocity to be mildly
relativistic based on the lack of the observed counter jet being explained by
the relativistic Doppler beaming. The results of S01 combined with the
constraints of Gleissner et al.\ (2004) (based on the lack of radio--X-ray
correlations on m--h time scales) yield the jet velocity of $\simeq
(0.5$--$0.7) c$. This requires an acceleration $\ga 10^2$ times higher than
that estimated by us. We note, however, that our results do not constrain at
all the jet region at $z\ga 30 a$. Thus, the acceleration there can be much
higher, e.g., due to the stellar wind density being much lower than in the
inner jet region.

\subsection{Superorbital modulation}
\label{super}

Another argument for the relativistic bulk velocity of the jet electrons
emitting synchrotron radiation is the presence of the superorbital (i.e., on a
time scale much longer than the orbital period) modulation. The modulation,
with the period of $\simeq 150$ d, is present in both radio and X-rays (e.g.,
Brocksopp et al.\ 1999a; L06). The standard (though not proven) explanation of
that effect in Cyg X-1 (and a number of other X-ray binaries) is precession
(e.g., Larwood 1998; L06). Then, the jet precession can cause the flux
variability at this period if the jet is at least mildly relativistic. The
radio data on the superorbital modulation are compatible with this picture as
the modulation depth at the three monitored frequencies, 2.25, 8.3 and 15 GHz,
is compatible with being constant, and there are no significant phase lags. We
note, however, that those data do not require the jet velocity to be
relativistic in the core, as the superorbital modulation can be caused by
electrons outside it. Also, the required velocity can be as low as $\sim\! 0.1
c$ if the (unknown) precession amplitude is $\sim\! 30\degr$. 

On the other hand, we can provide here an explanation for the superorbital
radio modulation without invoking a relativistic jet velocity. Namely, a
precessing jet moves with respect to the wind structure (which is defined only
by the binary orbit and does not precess). The line of sight from a given point
in the jet, and thus the amount of the orbit-averaged free-free absorption in
the wind, will then change with the superorbital phase (without phase lags).
However, the modulation depth is frequency-dependent in this model. In
principle, it is possible to find parameters of the system yielding the same
modulation depth from 2 to 15 GHz, but it requires certain fine-tuning. 

\subsection{Two-flow jet}
\label{two_flow}

As we have discussed above, the observed phase lags of the orbital radio
modulation require that the bulk of the jet mass is nonrelativistic within the
region where most of the radio emission takes place. On the other hand, the
non-detection of the counter jet and the superorbital radio modulation point to
(though not completely require) relativistic bulk motion of most of the
synchrotron-emitting electrons. This apparent contradiction may be reconciled
provided that the jet contains two flows at different velocities, e.g., an
outer slow part and a fast inner part. 

Such a model for radio-emitting X-ray binaries has been recently proposed by
Ferreira et al.\ (2006). It is based on an analogous, two-flow, model for
extragalactic radio jets of Sol, Pelletier \& Ass\'eo (1989). The main
contribution to the total jet power is an electron-proton plasma that is either
nonrelativistic in X-ray binaries or mildly relativistic in AGN jets. This
plasma is generated by an MHD process in a large-scale magnetic field anchored
in the accretion disc. The second flow is composed of relativistic e$^\pm$
pairs, and it carries only a small fraction of the total power. It is produced
in the inner regions of the MHD flow, where the pairs are created and then
accelerated. The outer flow confines the inner one. In the AGN case, the outer
slow flow is identified with kpc-scale jets, fuelling the radio lobes and
hotspots, and  the fast inner flow is identified with pc-scale jets and the
superluminal motion (Sol et al.\ 1989). We note that the specific model of
Ferreira et al.\ (2006) postulates that the threshold for pair production,
forming the inner fast jet, is reached only above the accretion rates
corresponding to the hard spectral state.

A massive disk outflow (linked with the super-Eddington mass transfer rate) has
been proposed to exist in the X-ray binary SS~433 (Zitter, Calvani \&
D'Odoricio 1991; Fabrika 1993, 2004). The outflow, with the velocity of $\sim
1.5 \times 10^8$ cm s$^{-1}$, is radiatively driven from outer parts of the
accretion disc. Its direction is determined by the tilt of the disc and is
subject to both precessional and nutational motions of the outer disc.
Begelman, King \& Pringle (2006) then postulated that the outflow is massive
enough to deflect the relativistic jet launched from the vicinity of the
accreting object and to align it with the outflow axis. 

In the case of Cyg X-1, we may identify the slow flow with the one needed to
explain the radio phase lags, and the fast one, with that at a relativistic
motion explaining the invisibility of the counter jet and the superorbital
modulation. However, this identification requires modifications of the above
models to allow for the existence of the two flows already at lower accretion
rates, corresponding to the hard spectral state.

We further note that our postulated slow (and dark) flow is very likely to be
identical to that found by Gallo et al.\ (2005) to be required to power the
radio lobes of Cyg X-1. The kinetic power of that flow was found by Gallo et
al.\ (2005) to be comparable to the X-ray luminosity. On the other hand, Heinz
(2006) found that the kinetic power carried by fields and radio-emitting
electrons in the visible jet of Cyg X-1 was much smaller. This is in agreement
with the picture in which the visible (fast) flow is confined by the dark
(slow) one. Also, Heinz (2006) estimated the velocity of the dark flow powering
the radio lobes of Cyg X-1 as $\ga 6 \times 10^8$ cm s$^{-1}$. This is in
agreement with the velocity constrained by the radio modulation, see Section
\ref{general} above. 

\subsection{Cyg X-3} 
\label{cygx3}

Cyg X-3 is an example of another high-mass X-ray binary with the companion
emitting strong wind and with very strong orbital modulation of X-rays, thought
to be caused by absorption and scattering in the wind. It also has strong radio
emission, most likely originating from a jet of that system. Still, no orbital
modulation of the radio emission is seen (Hjalmarsdotter et al.\ 2004).

To explain this lack of modulation, we first point out that Cyg X-3 is a much
more compact system than Cyg X-1. The orbital period of the former is only 4.8
hr. Cherepashchuk \& Moffat (1994) have estimated its orbital separation as
$a\simeq (2.2$--$3.9) \times 10^{11}$ cm, an order of magnitude less than
that of Cyg X-1. The degree of asymmetry of the wind at a given height, $z$, is
proportional to $a/z$. Thus, if the jet of Cyg X-3 emits a given radio
frequency at the same height as Cyg X-1, its orbital modulation will be still
much lower. 

Furthermore, the companion star in Cyg X-3 is a Wolf-Rayet star, which mass
loss rate, estimated as $-\dot M\simeq (3.6$--$29) \times 10^{-5} \msun$
yr$^{-1}$ (Ogley, Bell Burnell \& Fender 2001), is much higher than that of the
OB star in Cyg X-1. Thus, the radio emission originating relatively close to
the jet base is likely to be completely free-free absorbed in Cyg X-3. The high
mass-loss rate coupled with the small separation also makes the wind optically
thick to X-rays, not allowing us to use our optically-thin model (Section
\ref{wind}). 

The height from which a radio frequency is predominantly observed is then
roughly at $z \propto L_{\rm J}^{2/3}$ (Blandford \& K\"onigl 1979). The power
of the jet is, in turn, $L_{\rm J}\propto L_{\rm X}^p$, with $p\simeq 0.5$--1
(e.g., Migliari \& Fender 2006). The X-ray luminosity of Cyg X-3 is $\sim\!
(10$--100) times that of Cyg X-1 (e.g., Gallo, Fender \& Pooley 2003; Vilhu et
al.\ 2003). This consideration also points out to the height of radio emission
in Cyg X-3 being significantly higher than that in Cyg X-1. Consequently, the
wind at the radio-emitting region in Cyg X-3 is likely to be almost completely
axially symmetric, resulting in no orbital modulation in the radio range. 

\section{Conclusions}
\label{conclusions}

Our main conclusions are as follows. We find the observational data on the
orbital modulation of the radio emission from Cyg X-1 fully consistent with the
emission originating in a jet. As postulated before, the modulation is caused
by free-free absorption in the wind from the companion, with the line of sight
changing during the orbital motion. 

We find a crucial role of X-ray irradiation of the wind in enhancing its axial
asymmetry with respect to the jet axis. The irradiation strongly increases the
wind temperature, up to the Compton temperature of $\sim\! 10^6$ K, in the
vicinity of the X-ray source, which then strongly reduces the free-free
absorption on the side of the black hole. It also increases the overall wind
temperature to $\ga 3\times 10^5$ K (except for the vicinity of the companion),
which is about one order of magnitude above the star temperature. This reduces
the constant component of the free-free absorption and allows the transmitted
fraction of the radio emission to be close to unity. Consequently, the
intrinsic jet emission is likely to be close to the observed flat radio
spectrum (Fender et al.\ 2000). The wind temperatures calculated by us agree
with those inferred from the {\it Chandra\/} data (Miller et al.\ 2005). 

Another crucial requirement following from our study is the velocity of the
bulk of the mass of the jet being nonrelativistic, $\sim\! 5 \times 10^8 $ cm
s$^{-1}$, in the jet core (within $\sim\! 10^{14}$ cm). It follows from the
relatively large phase lags of the absorption maxima with respect to the
spectroscopic zero phase. The only explanation of these lags appears to be a
kinematic bending of the jet caused by the initial jet velocity in the binary
plane (equal to the orbital velocity of the black hole) being comparable
(though still substantially lower) to the jet vertical velocity in the region
where most of the 2--15 GHz emission is produced. Also, the data imply that the
above velocity is reached following an approximately constant acceleration
within the core, while a rapid, instantaneous-like acceleration is ruled out. 

However, the jet may then become mildly relativistic outside of the core, as
implied by the lack of a detectable counter jet (S01). Also, the
synchrotron-emitting electrons may have a relativistic bulk motion already in
the core if the jet is composed of two components, a slow, heavy, dark outer
flow, and a relativistic, light, and bright inner flow (as proposed by several
authors in order to explain other observational features of jets). 

We also consider the observed superorbital modulation of the radio emission,
with the period of $\sim\! 150$ d. That modulation can be caused by the Doppler
effect in a relativistic precessing jet. However, we find we can also explain
that modulation by the free-free absorption varying due to the jet precession.
When the precessing jet is inclined towards the observer, the average
absorption of its emission is weaker.  

We construct specific models constraining the distribution of the radio
emission at a given frequency along the jet. The lower the frequency, the
higher up it is emitted, and thus it is less modulated by the absorption, as
observed. Within the 90 per cent confidence intervals, our results are
compatible with the $z\propto 1/\nu$ law of Blandford \& K\"onigl (1979). Our
results are also compatible with the plausible values of the inclination of Cyg
X-1, $\sim\! 30\degr$--$60\degr$. For a given model, the higher the
inclination, the deeper the modulation. 

A prediction from our models is that the modulation depth should increase with
frequency also at $\nu> 15$ GHz. Thus, future monitoring at those frequencies
would be of great importance for our understanding of the jet structure.

We also consider the issue why there is no observed radio modulation in Cyg
X-3, which system is also powered by wind accretion and which shows strong
orbital modulation of its X-ray emission. We find a major reason for the lack
of radio modulation is its orbital separation being an order of magnitude lower
than in Cyg X-1, thus strongly reducing the wind asymmetry seen from the jet.
Also, it has much higher both the wind rate and the luminosity, which both
factors imply radio emission take place at higher heights than in Cyg X-1
(further reducing any asymmetry of the wind seen from the jet). 

Finally, we point out that our treatment of the jet and the wind has still
neglected their dynamical interactions. We also treat jet emission only in the
core (where it is modulated), and do not model its observed extended component.
These issues can be addressed in future work. 

\section*{Acknowledgments}
We thank M. Sikora for valuable discussions, R. Goosmann, J. Poutanen and the
referee for useful comments, C. Brocksopp for discussion of her results and the
numerical code used in BFP02, and P. Lachowicz for providing us with the data
shown in fig.\ 4 of L06. This research has been supported by the grants
1P03D01827, 1P03D01128 and 4T12E04727.

\bsp

\label{lastpage}

\end{document}